\documentclass[superscriptaddress,reprint, prl]{revtex4-1}
\usepackage{graphicx}
\usepackage{amsmath}
\usepackage{amssymb}
\usepackage[squaren, thinqspace]{SIunits}
\usepackage{xspace}
\usepackage{setspace}
\usepackage{color}
\begin{document}

\newcommand{\CFS}{\mathrm{Co}_{2}\mathrm{FeSi}}
\newcommand{\CFA}{\mathrm{Co}_{2}\mathrm{FeAl}}
\newcommand{\magnetite}{\mathrm{Fe}_{3}\mathrm{O}_{4}}
\newcommand{\gSpinMix}{g_{\uparrow\!\downarrow}}

\newcommand{\stbg}{\textcolor{black}}
\newcommand{\weg}{\sout}

\title{Scaling behavior of the spin pumping effect in ferromagnet/platinum bilayers}

\author{F.\,D.~Czeschka}
 \affiliation{Walther-Mei{\ss}ner-Institut, Bayerische Akademie der Wissenschaften, 85748 Garching, Germany}

\author{L.~Dreher}
\author{M.\,S.~Brandt}
 \affiliation{Walter Schottky Institut, Technische Universit\"{a}t M\"{u}nchen, 85748 Garching, Germany}

\author{M.~Weiler}
 \author{M.~Althammer}
 \affiliation{Walther-Mei{\ss}ner-Institut, Bayerische Akademie der Wissenschaften, 85748 Garching, Germany}

\author{I.-M.~Imort}
\author{G.~Reiss}
\author{A.~Thomas}
 \affiliation{Fakult\"{a}t f\"{u}r Physik, Universit\"{a}t Bielefeld, 33602 Bielefeld, Germany}

\author{W.~Schoch}
\author{W.~Limmer}
 \affiliation{Institut f\"{u}r Quantenmaterie, Universit\"{a}t Ulm, 89069 Ulm, Germany}

\author{H.~Huebl}
 \affiliation{Walther-Mei{\ss}ner-Institut, Bayerische Akademie der Wissenschaften, 85748 Garching, Germany}

\author{R.~Gross}
 \affiliation{Walther-Mei{\ss}ner-Institut, Bayerische Akademie der Wissenschaften, 85748 Garching, Germany}
 \affiliation{Physik-Department, Technische Universit\"{a}t M\"{u}nchen, 85748 Garching, Germany}

\author{S.\,T.\,B.~Goennenwein}
 \email{goennenwein@wmi.badw.de}
 \affiliation{Walther-Mei{\ss}ner-Institut, Bayerische Akademie der Wissenschaften, 85748 Garching, Germany}

\begin{abstract}
We systematically measured the DC voltage $V_\mathrm{ISH}$ induced by spin pumping together with the inverse spin Hall effect in ferromagnet/platinum bilayer films. In all our samples, comprising  ferromagnetic $3d$ transition metals, Heusler compounds, ferrite spinel oxides, and magnetic semiconductors, $V_\mathrm{ISH}$ invariably has the same polarity\stbg{, and} scales with the magnetization precession cone angle. These findings\stbg{, together with the spin mixing conductance derived from the experimental data,} quantitatively corroborate the present theoretical understanding of spin pumping in combination with the inverse spin Hall effect.
\end{abstract}

\maketitle

Spin current related phenomena are an important aspect of modern magnetism\stbg{~\cite{ralph_spin_2008,takahashi_spin_2008,zutic_spintronics:_2004,maekawa_concepts_2006}}. For example, pure spin currents -- a directed flow of angular momentum without an accompanying net charge current -- can propagate in magnetic insulators~\cite{kajiwara_transmission_2010}. A spin current $\mathbf{J}_{\mathrm{s}}$ can be detected via the inverse spin Hall (ISH) effect, where $\mathbf{J}_{\mathrm{s}}$ with polarization orientation $\mathbf{\hat{s}}$ is converted into a charge current $\mathbf{J}_{\mathrm{c}}\propto \alpha_{\mathrm{SH}} \left(\mathbf{\hat{s}} \times \mathbf{J}_{\mathrm{s}}\right)$ perpendicular to both $\mathbf{\hat{s}}$ and $\mathbf{J}_{\mathrm{s}}$~\cite{hirsch_spin_1999,takahashi_spin_2008}.
Recently, Mosendz \textit{et al.}~\cite{mosendz_quantifying_2010} showed that the spin Hall angle $\alpha_{\mathrm{SH}}$ can be quantitatively determined from so-called spin pumping experiments in ferromagnet/normal metal (F/N) bilayers. Here, the magnetization of the F layer is driven into ferromagnetic resonance (FMR) and can relax by emitting a spin current into the adjacent N layer~\cite{tserkovnyak_spin_2002,saitoh_conversion_2006}. Spin pumping can thus be understood as the inverse of spin torque\stbg{~\cite{ralph_spin_2008,katine_current-driven_2000}}, and gives access to the physics of spin currents, magnetization dynamics, and damping. The present theoretical models~\cite{tserkovnyak_enhanced_2002,tserkovnyak_spin_2002} suggest that spin pumping in conductive ferromagnets is a generic phenomenon, where the magnitude of $\mathbf{J}_{\mathrm{s}}$ is governed by the magnetization precession cone angle and the spin mixing conductance $\gSpinMix$ at the F/N interface. However, most spin pumping experiments to date have been performed in transition metals~\cite{urban_gilbert_2001,saitoh_conversion_2006,costache_electrical_2006}, so that generic properties could not be addressed.
In this letter, we provide experimental evidence that the present theories for spin pumping are not limited to transition metal-based bilayers, but also apply to the ferromagnetic Heusler compounds $\CFA$ and $\CFS$, the ferrimagnetic oxide spinel $\magnetite$, and the dilute magnetic semiconductor (Ga,Mn)As (DMS). We demonstrate this by simultaneous DC voltage and FMR measurements that yield the correlation of the inverse spin Hall voltage $V_\mathrm{ISH}\propto J_{\mathrm{c}}$ along $\mathbf{J}_{\mathrm{c}}$ and the magnetization precession cone angle $\Theta$ in FMR.  Our experimental findings clearly confirm the scaling behavior of $V_\mathrm{ISH}$ suggested by theory.

\begin{figure}
  \includegraphics[width=8cm]{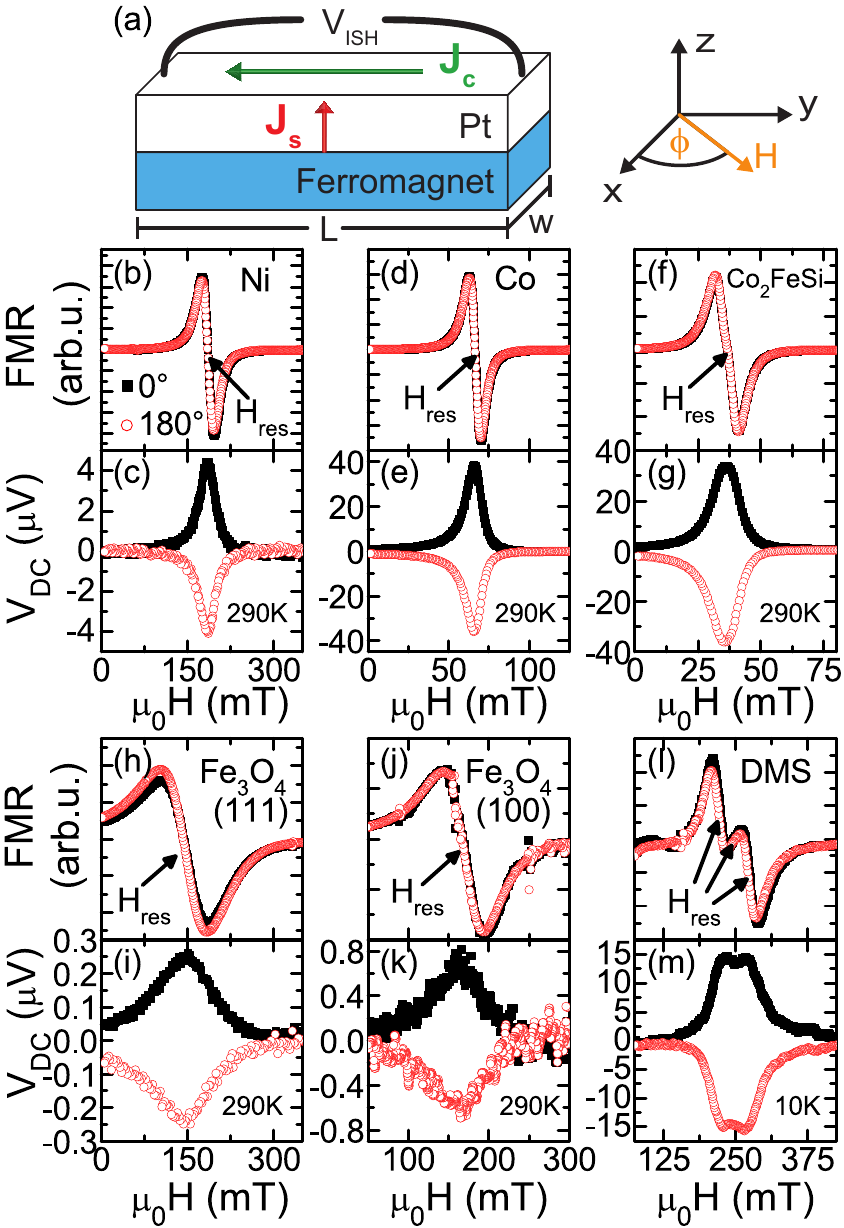} 
\caption{(a) Sketch of the the coordinate system and the F/Pt bilayer sample. (b),(d),(f),(h),(j),(l) show the FMR signal of F/Pt bilayers, with F as quoted in the individual panels, recorded with $\mathbf{H}$ parallel (black full squares) and antiparallel (red open circles) to $\mathbf{\hat{x}}$. DMS stands for $200\,\mathrm{nm}$ (Ga,Mn)As. (c),(e),(g),(i),(k),(m) show the DC voltage measured simultaneously with the respective FMR traces.}
\label{fig:Materials}
\end{figure}

We fabricated F/Pt bilayers, using Ni, Co, Fe, $\CFA$, $\CFS$, $\magnetite$, and (Ga,Mn)As, for the F layer. The Ni, Co and Fe films were deposited on oxidized silicon substrates via electron beam evaporation at a base pressure of  $1\times10^{-8}\,\mathrm{mbar}$. The Heusler compounds were sputtered on (001)-oriented MgO single crystal substrates at an Ar pressure of $1.5\times10^{-3}\,\mathrm{mbar}$, followed by annealing  at $500\,\degreecelsius$~\cite{ebke_low_2010}. Epitaxial (100)- and (111)-oriented $\magnetite$ films were grown via pulsed laser deposition in argon atmosphere on (100)-oriented MgO and (0001)-oriented  Al$_{2}$O$_{3}$ substrates, respectively, at a substrate temperature of $320\,\degreecelsius$~\cite{venkateshvaran_epitaxial_2009}. The Ga$_{1-x}$Mn$_x$As ($x=0.04$) films were grown via low temperature molecular beam epitaxy on (001)-oriented GaAs substrates~\cite{limmer_angle-dependent_2006}. All F layers have a thickness $t_{\mathrm{F}}=10\,\mathrm{nm}$, except for the (111)-oriented $\magnetite$ film, which has $t_{\mathrm{F}}=35\,\mathrm{nm}$, and the (Ga,Mn)As films with $t_{\mathrm{F}}=200\,\mathrm{nm}$, $175\,\mathrm{nm}$  and $65\,\mathrm{nm}$.
As a high-quality, transparent interface is crucial for spin pumping~\cite{mosendz_suppression_2010}, all F layers were covered in situ with $t_{\mathrm{N}}=7\,\mathrm{nm}$ of Pt, except for the (Ga,Mn)As films, which were covered after exposure to ambient atmosphere. All samples were cut into rectangular bars (length $L=3\,\mathrm{mm}$, width $w=1\,\mathrm{mm}$  or $2\,\mathrm{mm}$) and contacted on the short sides for electrical measurements as shown in Fig.~\ref{fig:Materials}(a).

The FMR and spin pumping experiments were performed in a magnetic resonance spectrometer at a fixed microwave frequency $\nu_\mathrm{MW}=9.3\,\mathrm{GHz}$ as a function of an externally applied static magnetic field $\mathbf{H}$, in the temperature range from $2\,\kelvin$ to $290\,\kelvin$. We took care to position the respective sample on the axis of the $\mathrm{TE}_{102}$ microwave cavity, in order to locate it in an antinode of the microwave magnetic field and in a node of the microwave electric field. The FMR was recorded using magnetic field modulation and lock-in detection, so that the resonance field $H_{\mathrm{res}}$ corresponds to the inflection point in the FMR spectra. The DC voltage $V_{\mathrm{DC}}$ between the contacts indicated in Fig.~\ref{fig:Materials}(a) was measured with a nanovoltmeter.

\begin{figure}
\includegraphics[width=8cm]{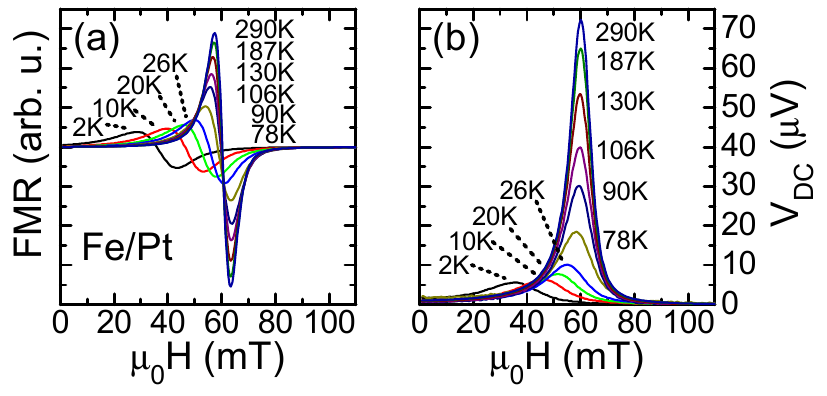}
\caption{Temperature-dependent evolution of (a) the FMR, and (b) the $V_{\mathrm{DC}}$ spectra of a Fe/Pt bilayer.}
\label{fig:T-dep}
\end{figure}

Figure~\ref{fig:Materials} shows a selection of FMR and $V_{\mathrm{DC}}$ spectra, recorded for two magnetic field orientations in the film plane: $\phi=0\degree$ corresponds to $\mathbf{H}$ parallel to $\mathbf{\hat{x}}$ (black full squares), while for $\phi=180\degree$, $\mathbf{H}$ is antiparallel to $\mathbf{\hat{x}}$ (red open circles). All measurements in Fig.~\ref{fig:Materials} were taken at $290\,\kelvin$, except for the (Ga,Mn)As data recorded at $T=10\,\kelvin$ (Fig.\,\ref{fig:Materials}(l) and (m)). Since the FMR is invariant with respect to magnetic field inversion, the FMR traces for $\phi=0\degree$ and $\phi=180\degree$ should superimpose, as indeed observed in experiment. The FMR signal of all samples in Fig.\,\ref{fig:Materials} consists of a single resonance line, with the exception of (Ga,Mn)As, in which several standing spin wave modes contribute to the FMR spectrum \cite{bihler_spin-wave_2009}. The $V_{\mathrm{DC}}$ traces show one clear extremum in $V_{\mathrm{DC}}$ at $H_{\mathrm{res}}$; only in (Ga,Mn)As, several $V_{\mathrm{DC}}$ extrema corresponding to the spin wave modes can be discerned. The magnitude of $V_{\mathrm{DC}}$ ranges from a few $100\,\nano\volt$ in $\magnetite$ to a few 10 $\micro\volt$ in $\CFS$ and Fe (Fig.~\ref{fig:T-dep}). In contrast to the FMR, the $V_{\mathrm{DC}}$ extremum changes sign when the magnetic field is reversed. It also is important to note that $V_{\mathrm{DC}}$ always has a maximum ($V_{\mathrm{DC}}>0$) for $\phi=0\degree$, whereas a minimum ($V_{\mathrm{DC}}<0$) is observed for $\phi=180\degree$.

We furthermore studied the evolution of the FMR and $V_{\mathrm{DC}}$ signals as a function of temperature in several bilayer samples. As an example, the FMR and $V_{\mathrm{DC}}$ spectra of an Fe/Pt sample, recorded for a series of temperatures $2\,\kelvin\leq T \leq 290\,\kelvin$ are shown in Fig.\,\ref{fig:T-dep}. With decreasing $T$, the FMR broadens and shifts to lower $H_{\mathrm{res}}$, as does the peak in $V_{\mathrm{DC}}$.

We now turn to the interpretation of the experimental data of Figs.\,\ref{fig:Materials} and \ref{fig:T-dep}. We attribute the peaks in $V_{\mathrm{DC}}$ to spin pumping in combination with the inverse spin Hall effect in the F/Pt bilayers ~\cite{mosendz_quantifying_2010,saitoh_conversion_2006,tserkovnyak_spin_2002}. This naturally explains that  $V_{\mathrm{DC}}$ changes sign when the $\mathbf{H}$ orientation is inverted from $\phi$=0$\degree$ to 180$\degree$, as $\mathbf{H}$ determines the orientation of the spin polarization vector $\mathbf{\hat{s}}$ in $\mathbf{J}_{\mathrm{c}}\propto \left(\mathbf{\hat{s}} \times \mathbf{J}_{\mathrm{s}}\right)$. Hence, $\mathbf{J}_{\mathrm{c}}$ and thus also $V_{\mathrm{DC}}$ is reversed if the magnetic field is inverted. \stbg{We note that the $H_{\mathrm{res}}$ are well above the coercive and the anisotropy fields of the respective ferromagnets, such that the magnetization $\mathbf{M}\parallel\mathbf{H}$ in good approximation.} Furthermore, the experimental observation that $V_{\mathrm{DC}}$ invariably has the same polarity for a given field orientation $\phi$, irrespective of the ferromagnetic material used in the F/Pt bilayer and of the measurement temperature, is fully consistent with spin pumping theory~\cite{tserkovnyak_enhanced_2002,tserkovnyak_spin_2002, tserkovnyak_nonlocal_2005, wang_voltage_2006, mosendz_quantifying_2010, mosendz_detection_2010}. We note that other mechanisms for the generation of a DC voltage in conjunction with FMR have been suggested and were observed in experiment~\cite{egan_dc_1963,gui_quantized_2007,mecking_microwave_2007}. Microwave rectification effects linked to the anisotropic magnetoresistance or the anomalous Hall effect often are superimposed onto the spin pumping signal, in particular if the sample is not located in a node of the microwave electric field~\cite{mosendz_quantifying_2010,inoue_detection_2007}. However, we rule out such mechanisms as the origin of the  $V_{\mathrm{DC}}$ observed in our experiments (Figs.\,\ref{fig:Materials} and \ref{fig:T-dep}) for two reasons. First, we have positioned the sample in a node of the microwave electric field, which minimizes rectification-type processes. Second, and more importantly, both the spontaneous resistivity anisotropy $\Delta \rho$ determining the anisotropic magnetoresistance and the anomalous Hall coefficient $R_\mathrm{H}$ are substantially different in magnitude and in sign for the different ferromagnetic materials in our F/Pt bilayers~\cite{Ohandley:ModernMagneticMaterials}. Nevertheless, for a given $\phi$, we invariably observe the same $V_{\mathrm{DC}}$ polarity, which is difficult to rationalize for rectification effects.

\begin{figure}
 \includegraphics[width=7.5cm]{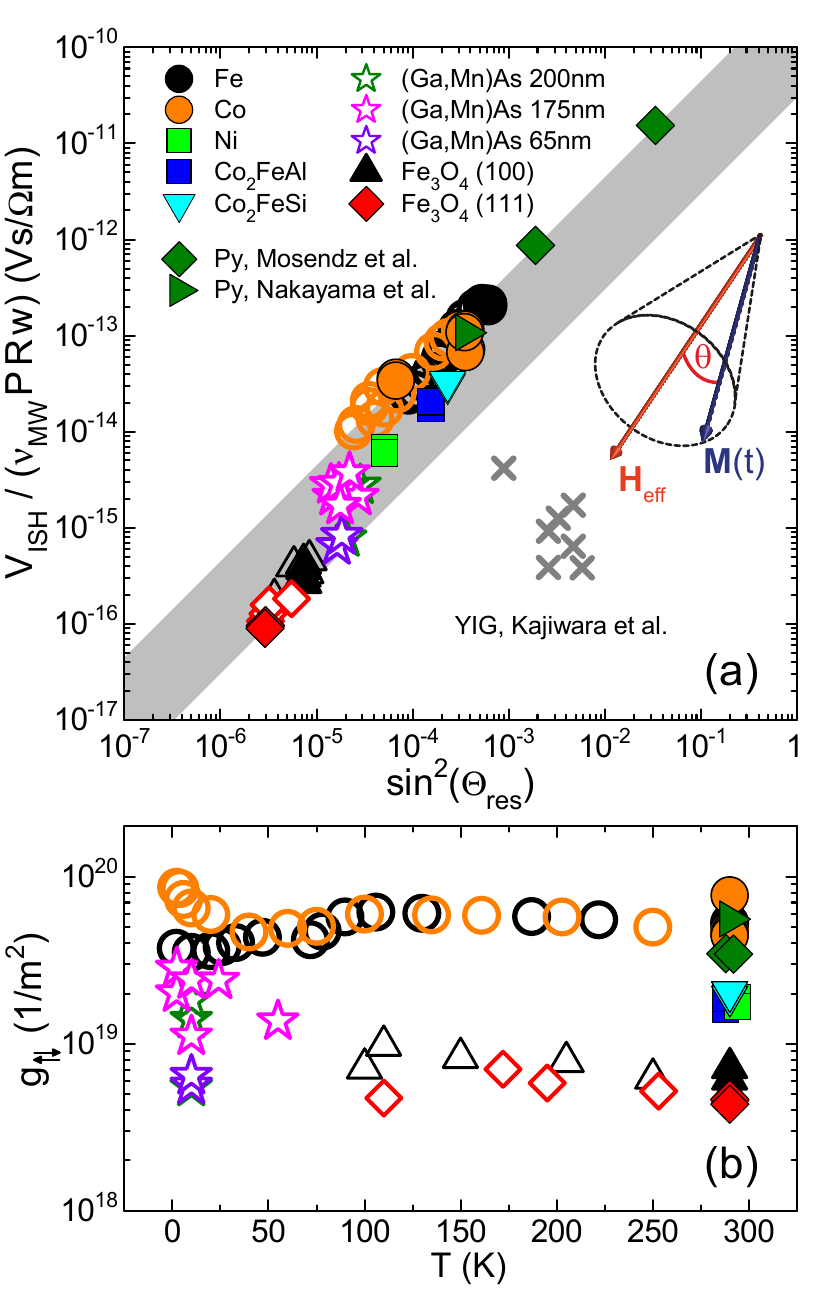}
\caption{(a) In all F/Pt bilayers made from conductive ferromagnets, the DC \stbg{voltage} $V_{\mathrm{ISH}}$ 
induced by a collective mode FMR scales with $\sin^{2}\Theta_{\mathrm{res}}$ \stbg{to within a factor of 10, as indicated by the grey bar.} The inset depicts the magnetization precession around the effective magnetic field. (b)  \stbg{From the scaling analysis, the spin mixing conductance $\gSpinMix$ can be  quantified as a function of temperature} (see text).  In panels (a) and (b), full symbols represent data taken at $290\,\kelvin$, open symbols correspond to data measured at lower $T$. \stbg{The Py and YIG data are taken from the literature, Refs.\,\cite{mosendz_detection_2010,mosendz_quantifying_2010,nakayama_inverse_2010,kajiwara_transmission_2010}.}}
\label{fig:scaling}
\end{figure}

In order to quantitatively compare our experimental data with spin pumping theory, we start from
\begin{equation}\label{eq:Mosendz:V-ISH}
V_{\mathrm{ISH}}
= \frac{-e \,\alpha_{\mathrm{SH}}\lambda_{\mathrm{SD}}\tanh\frac{t_{\mathrm{N}}}{2\lambda_{\mathrm{SD}}}}{\sigma_{\mathrm{F}}t_{\mathrm{F}}+\sigma_{\mathrm{N}}t_{\mathrm{N}}}\gSpinMix \nu_{\mathrm{MW}} L P \sin^{2}\Theta
\end{equation}
derived by Mosendz \textit{et al.}~\cite{mosendz_quantifying_2010,mosendz_detection_2010} for the inverse spin Hall DC voltage $V_{\mathrm{ISH}}$ arising due to spin pumping in permalloy/N bilayers\stbg{, assuming that the $N$ layer is an ideal spin current sink}. Here, $e$ is the electron charge, $\lambda_{\mathrm{SD}}$ is the spin diffusion length in N, $\gSpinMix$ is the effective spin mixing conductance~\cite{mosendz_detection_2010}, $\Theta$ is the magnetization precession cone angle (cf. inset in Fig.~\ref{fig:scaling}(a)), and $P$ is a correction factor taking into account the ellipticity of the magnetization precession~\cite{ando_optimum_2009, mosendz_detection_2010}. \stbg{For our samples, we calculated $0.5\le P\le 1.3$.} Note that Eq.\,\eqref{eq:Mosendz:V-ISH} has been adapted to our experimental configuration, and accounts for both the conductivity $\sigma_\mathrm{N}$ and $\sigma_\mathrm{F}$ of the N and F layer contributing to the bilayer conductivity.

Since \stbg{invariably} $t_{\mathrm{N}}=7\,\nano\meter$ and N=Pt for all our
bilayer samples,
$C\equiv\alpha_{\mathrm{SH}}\lambda_{\mathrm{SD}}\tanh(t_{\mathrm{N}}/2\lambda_{\mathrm{SD}})$
is a constant \stbg{at a given temperature}.
\stbg{In addition, the denominator in Eq.\,(\ref{eq:Mosendz:V-ISH}) can be expressed in terms of the sample geometry $w/L$ and resistance $R_{}$, measured in four point experiments: $\sigma_{\mathrm{F}}t_{\mathrm{F}}+\sigma_{\mathrm{N}}t_{\mathrm{N}}=(R_{} w/L)^{-1}$. We thus rewrite Eq.\,(\ref{eq:Mosendz:V-ISH}) as
\begin{equation}\label{eq:Mosendz:V-ISH:simplified}
\frac{V_{\mathrm{ISH}} }{\nu_{\mathrm{MW}} P R\, w} =-e C \gSpinMix \sin^{2}\Theta\,.
\end{equation}
The} theoretical models for the spin mixing
conductance~\cite{brataas_finite_2000,tserkovnyak_enhanced_2002}
suggest that $\gSpinMix$ of conductive ferromagnet/normal metal
interfaces is determined mainly by the N layer, i.e., the Pt
layer in our case. \stbg{In other words, } $\gSpinMix$ should be of comparable
magnitude in all our samples. \stbg{Equation (\ref{eq:Mosendz:V-ISH:simplified}) then represents a scaling relation for }
all F/Pt bilayers made from \stbg{a} conductive ferromagnet \stbg{and a Pt layer of one and the same thickness $t_{\mathrm{N}}$, irrespective}
of the \stbg{particular} ferromagnetic material, its
magnetic properties, or the details of the charge transport
mechanism such as band conduction or charge carrier hopping.

We now test the scaling relation of Eq.\,\eqref{eq:Mosendz:V-ISH:simplified} against our experimental
data.  At ferromagnetic resonance, the magnetization precession
cone angle is $\Theta_{\mathrm{res}}=2 h_{\mathrm{MW}}/
(\sqrt{3}\Delta  H_{\mathrm{pp}} )$~\cite{guan_phase_2007},
with the microwave magnetic field
$h_{\mathrm{MW}}=0.12\,\mathrm{mT}$   as determined in
paramagnetic resonance calibration experiments. We extract the
FMR peak-to-peak line width $\Delta H_{\mathrm{pp}}$ from the
experimental data, and use the measured DC voltage
$V_{\mathrm{DC,res}}$ at $H_\mathrm{res}$ to determine
\stbg{$V_{\mathrm{ISH}}=V_{\mathrm{DC,res}}$}. Figure~\ref{fig:scaling}(a)
shows \stbg{$V_{\mathrm{ISH}}/\left(\nu_{\mathrm{MW}} P R\,w\right)$} versus
$\sin^{2}\Theta_{\mathrm{res}}$ thus obtained. Full symbols
indicate data measured at $290\,\kelvin$, while measurements at
lower temperatures are shown as open symbols. Data for
permalloy/Pt (Py/Pt) extracted from
Refs.~\cite{mosendz_quantifying_2010,mosendz_detection_2010,nakayama_inverse_2010}
are also included in the figure. \stbg{For the sake of
completeness, data for
$\mathrm{Y}_{3}\mathrm{Fe}_{5}\mathrm{O}_{12}$/Pt (YIG/Pt), taken
from Ref.~\cite{kajiwara_transmission_2010}, are also shown. Since YIG is an insulator, however,
$\gSpinMix$ is dominated by its imaginary part, in contrast to
the mostly real $\gSpinMix$ for conductive
ferromagnets~\cite{xia_spin_2002,carva_spin-mixing_2007,tserkovnyak_enhanced_2002}.
Moreover, spin wave modes govern the YIG FMR signal, impeding a
straightforward analysis~\cite{sandweg_enhancement_2010}. Thus, we here 
focus only on conductive ferromagnet/Pt bilayers. In these samples, $V_{\mathrm{ISH}}/\left(\nu_{\mathrm{MW}} P R\,w\right)$ indeed scales as suggested by Eq.\,\eqref{eq:Mosendz:V-ISH:simplified} to within a factor of 10 (grey bar in Fig.\,\ref{fig:scaling}(a)). The deviations from perfect scaling are due to a slight material dependence of $\gSpinMix$, as detailed in the next paragraph. The scaling} behavior is observed over more than four orders of
magnitude in \stbg{$V_{\mathrm{ISH}}/\left(\nu_{\mathrm{MW}} P R\,w\right)$}  and
$\sin^{2}\Theta_{\mathrm{res}}$, for samples made from
\stbg{conductive} ferromagnetic films with qualitatively different exchange
mechanisms, \stbg{transport} properties, crystalline quality, and
crystalline structure. Moreover, F/N bilayer samples fabricated and investigated by
different groups are consistently described. 


To quantify $\gSpinMix$ of a given bilayer, \stbg{we write Eq.\,(\ref{eq:Mosendz:V-ISH:simplified}) as} $\gSpinMix = - V_{\mathrm{ISH}}/ \left[\nu_{\mathrm{MW}} P R\,w e C \sin^{2}\Theta_{\mathrm{res}}\right]$. {\stbg U}sing the room temperature values $\alpha_{\mathrm{SH}}=0.013$ and $\lambda_{\mathrm{SD}}=10\,\nano\meter$ for Pt~\cite{mosendz_detection_2010,vila_evolution_2007}, $P$ calculated as detailed in Ref.~\cite{mosendz_detection_2010} (and literature values for the conductivities of Py and Pt \cite{ando_electric_2008,mosendz_quantifying_2010} for the data points extracted from Refs.~\cite{mosendz_quantifying_2010,mosendz_detection_2010,nakayama_inverse_2010})\stbg{, we obtain $\gSpinMix$ as} shown in Fig.\,\ref{fig:scaling}(b). Clearly, the conjecture that  $\gSpinMix$ is independent of the F layer properties is well fulfilled for highly conductive (``metallic'') ferromagnets, such as the $3d$ transition metals, permalloy, or the Heusler compounds, which all are in the range $\gSpinMix=\left(4\pm3 \right)\times 10^{19}\,\meter^{-2}$. In the low-conductivity ferromagnet $\magnetite$, $\gSpinMix$ is about a factor of 6 smaller\stbg{, but the linear scaling is still observed}. In (Ga,Mn)As, $\gSpinMix$ appears to be in between these two regimes. However, several spin wave modes contribute to the FMR \stbg{in (Ga,Mn)As} (cf. Fig.~\ref{fig:Materials}(l)) and a fit with at least three Lorentzian lines was required to reproduce the FMR and $V_{\mathrm{DC}}$ data. So the assumption of a single, position-independent magnetization precession cone angle $\Theta_{\mathrm{res}}$ is not warranted~\cite{bihler_spin-wave_2009}. \stbg{For standing spin waves, the magnetization precession amplitude $\Theta_{\mathrm{res}}(z)$ changes as a function of $z$ across the film thickness, which in turn can qualitatively alter the magnitude of }$V_{\mathrm{DC}}$~\cite{sandweg_enhancement_2010}. \stbg{Since $\Theta_{\mathrm{res}}(z)$ moreover depends on the particular spin wave mode excited, a} more thorough study of spin pumping due to spin wave modes is mandatory to evaluate $\gSpinMix$ in (Ga,Mn)As/Pt\stbg{. In addition, in systems with large spin-orbit coupling such as (Ga,Mn)As, the magnetization precession can even induce a charge current via an inverse, spin-orbit driven spin torque effect}~\cite{Hals_GaMnAs_2010}. \stbg{Taken together, the experimental results summarized in Fig.\,\ref{fig:scaling}(b) represent an incentive to theory to calculate $\gSpinMix$ for ferromagnets with different conductivity magnitude, transport mechanisms, and inhomogeneous spin texture.}

\stbg{Another interesting experimental observation from Fig.\,\ref{fig:scaling}(b) is that temperature has little influence on $\gSpinMix$. According to the present theoretical understanding, $\gSpinMix$ in diffusive bilayers is governed by the conductivity $\sigma_{\mathrm{N}}(T)$ of the normal metal \cite{tserkovnyak_enhanced_2002,brataas_finite_2000}. The weak temperature dependence of $\gSpinMix$ (Fig.\,\ref{fig:scaling}(b)) thus suggests that $\sigma_{\mathrm{N}}(T)$ of our Pt films also should not substantially change with temperature. This is corroborated by resistance measurements, which show that $\sigma_{\mathrm{N}}(T)$ increases by less than a factor of 2 from $290\,\kelvin$ to $2\,\kelvin$. Since $\alpha_{\mathrm{SH}}\propto\sigma^{0.6\ldots1}$~\cite{nakayama_detection_2010,vila_evolution_2007} is governed by $\sigma_{\mathrm{N}}(T)$, and since $\lambda_{\mathrm{SD}}$ in Pt increases by less than 50\% from $290\,\kelvin$ to $2\,\kelvin$~\cite{vila_evolution_2007}, $C=\alpha_{\mathrm{SH}}\lambda_{\mathrm{SD}}\tanh(t_{\mathrm{N}}/2\lambda_{\mathrm{SD}})$ changes by at most a factor of 3 in the whole temperature range investigated experimentally. This warrants the use of $C \gSpinMix$ as an essentially temperature-independent scaling constant in Eq.\,(\ref{eq:Mosendz:V-ISH:simplified}).}

%

In summary, we have measured the DC voltage caused by spin pumping and the inverse spin Hall effect in F/Pt samples, with F made from elemental $3d$ ferromagnets, the ferromagnetic Heusler compounds $\CFA$ and $\CFS$, the ferrimagnetic oxide spinel $\magnetite$, and the magnetic semiconductor (Ga,Mn)As. Although the magnetic exchange mechanism, the saturation magnetization, the spin polarization, the charge carrier transport mechanism and the charge carrier polarity are qualitatively different in the different samples, the DC voltage has identical polarity for all bilayers investigated, and its magnitude is well described by a scaling relation of the form of Eq.\,(\ref{eq:Mosendz:V-ISH:simplified}) within the entire temperature range $2\,\kelvin\le T\le 290\,\kelvin$ studied. Our experimental findings thus quantitatively corroborate the present spin pumping/inverse spin Hall theories~\cite{tserkovnyak_enhanced_2002,tserkovnyak_spin_2002,tserkovnyak_nonlocal_2005,wang_voltage_2006, mosendz_quantifying_2010,mosendz_detection_2010}, \stbg{and are an incentive for quantitative calculations of $\gSpinMix(T)$ in various types of F/N bilayers.}

We thank G.\,E.\,W. Bauer for valuable discussions. The work at WMI was financially supported via the Excellence Cluster ``Nanosystems Initiative Munich (NIM)''. A.T. and I.-M. I. are supported by the NRW MIWF.


\end{document}